\documentclass[12pt,a4paper]{article}

\catcode`\@=11
\@addtoreset{equation}{section}

\global\arraycolsep=2pt
\usepackage{mathrsfs,amsbsy,amssymb,latexsym,amsfonts,amsmath}
\usepackage{graphicx,color,cite}

\allowdisplaybreaks

\newcommand{\dd}{{\rm d}}
\newcommand{\pd}{{\cal D} }
\newcommand{\ds}{{\rm dS} }
\newcommand{\ads}{{\rm AdS} }

\usepackage[hypertex]{hyperref}

\begin{document}

\begin{flushright}
\parbox{4.2cm}
{KUNS-2539}
\end{flushright}

\vspace*{2cm}

\begin{center}
{\Large \bf Comments on Entanglement Entropy \\ in the dS/CFT Correspondence}
\vspace*{2cm}\\
{\large Yoshiki Sato\footnote{E-mail:~yoshiki@gauge.scphys.kyoto-u.ac.jp} 
}
\end{center}

\vspace*{1cm}
\begin{center}
{\it Department of Physics, Kyoto University \\ 
Kyoto 606-8502, Japan} 
\end{center}

\vspace{1cm}

\begin{abstract}
We consider the entanglement entropy in the dS/CFT correspondence.
In Einstein gravity on de Sitter spacetime we propose the holographic entanglement entropy as the 
analytic continuation of the extremal surface in Euclidean anti-de Sitter spacetime.
Even though dual conformal field theories for Einstein gravity on de Sitter spacetime are not  known yet,
we analyze the free $Sp(N)$ model dual to Vasiliev's higher spin gauge theory as a toy model.
In this model we confirm the behaviour similar to our holographic result from Einstein gravity.
\end{abstract}

\thispagestyle{empty}
\setcounter{page}{0}

\newpage

\section{Introduction}

The AdS/CFT correspondence  provides a remarkable connection between gravitational theories in 
anti-de Sitter spacetime (AdS) and
nongravitational theories \cite{M,GKP,W}.
This enables us to analyze quantum gravitational theories by using nongravitational theories.

\medskip

A useful quantity to analyze gravitational theories is the holographic entanglement entropy proposed in
\cite{RT1,RT2}\footnote{The covariant generalisation was proposed in \cite{HRT}.}.
The holographic entanglement entropy contains information on 
gravitational theories \cite{Swingle1,Raamsdonk1}. For instance,  
Einstein's equation can be reproduced from the holographic entanglement entropy \cite{NNPT,Lash}.

\medskip

It is natural to apply the AdS/CFT correspondence to our Universe.
However, since it is known that our Universe is approximately de Sitter spacetime (dS), 
not AdS,
we cannot use the AdS/CFT correspondence to analyze our Universe.

\medskip

The dS/CFT correspondence has been proposed in \cite{Wds,Sds,nonG}.
These proposals were abstract, and concrete examples did not exist.
Recently, Anninos, Hartman and Strominger have proposed a concrete example 
of the dS/CFT correspondence based on Giombi-Klebanov-Polyakov-Yin duality 
(the duality between Vasiliev's four-dimensional higher-spin gauge theory on Euclidean AdS (EAdS) 
and the three-dimensional $O(N)$ vector model) \cite{AHS} (see also \cite{Anninos} for a review).
The authors showed that EAdS  and the $O(N)$ vector model are related to dS and the $Sp(N)$ vector model 
via analytic continuation, respectively. 
It follows that Vasiliev's higher-spin gauge theory on dS is the holographic dual of the Euclidean $Sp(N)$ vector model
which lives in $\mathcal{I}^+$ in dS.
We are now in a position to analyze the dS/CFT correspondence using the concrete example.

\medskip 

In this paper, we investigate the connection between bulk geometry and holographic entanglement entropy
in Einstein gravity on dS.
However, the notion of extremal surfaces whose boundaries sit on $\mathcal{I}^+$ is obscure.
If the surfaces were space-like, their area would be smaller and smaller as the surfaces approached null. 
If the surfaces were time-like, their area would be imaginary, and the surfaces would not be closed in general.  
We discuss this issue based on analytic continuation because 
the analytic continuation enables us to obtain surfaces in dS, which satisfy the equation of motion
obtained from the variation of the area functional.

\medskip

The organization of this paper is as follows. 
In section 2 we propose the holographic entanglement entropy formula for Einstein gravity on dS.
We find extremal surfaces in dS based on a double Wick rotation from EAdS in Poincar\'{e} coordinates.
We comment on extremal surfaces in more general set of asymptotically dS.
In section 3 we calculate the entanglement entropy in the free Euclidean $Sp(N)$ model. 
We compare this result with the result in section 2 and confirm that our proposal is sensible qualitatively.
Section 4 is devoted to a conclusion and discussion.

\section{Proposal for Holographic Entanglement Entropy in Einstein Gravity on dS}

It is known that a black hole's (BH) entropy is given by an event horizon area divided 
by four times Newton's constant.
The BH entropy formula holds not only in asymptotically flat or AdS but also in asymptotically dS.
The holographic entanglement entropy is a generalised quantity of the BH entropy \cite{LM}, and it is given by 
an area of a extremal surface divided by four times Newton's constant.
It is natural to expect that the Ryu-Takayanagi formula holds even in dS as the BH entropy formula.

\medskip

Thus our task is to find  ``extremal surfaces'' in Einstein gravity on dS.
However, the notion of the extremal surface is obscure  as noted in the Introduction.
Our proposal is that the extremal surfaces in dS are given by the analytic continuation of extremal surfaces in EAdS.
This proposal allows for extremal surfaces  which extend in complex-valued coordinate spacetime, 
and lets us find complex surfaces 
as extremal surfaces\footnote{
The extremal surfaces in our proposal are space-like.
However, it would be possible to regard time-like surfaces as the extremal surfaces.
The time-like surfaces might not be appropriate for the holographic entanglement entropy formula because 
 these surfaces are not closed in general. 
If we consider that we attach a half sphere to a half of dS so that it represents the Hartle-Hawking state,
the time-like surfaces would be closed.
In this case, the holographic entanglement entropy
would  become a sum of a pure real part and a pure imaginary part.
Since this result largely disagrees with our result in section 3, we will not consider this possibility in this paper.}. 
In the next section, we will check the consistency of our proposal. 

\medskip

The metric of dS in Poincar\'{e} coordinates is given by 
\begin{equation}
\dd s^2=\ell _{\ds}^2 \frac{-\dd \eta ^2+\sum_{i=1}^d\dd x_i^2}{\eta^2}
\label{poincareds}
\end{equation}
where $\eta$ is the conformal time, and $\ell _{\ds}$ is a dS radius.
By performing a double Wick rotation,
\begin{equation}
\eta \to iz\,, \quad 
\ell _{\ds} \to i \ell _{\ads}\,,
\label{wick}
\end{equation}
the metric \eqref{poincareds} becomes the metric in the Poincar\'{e} EAdS,
\begin{equation}
\dd s^2=\ell _{\ads}^2 \frac{\dd z ^2+\sum_{i=1}^d\dd x_i^2}{z^2}\,.
\label{poincareads}
\end{equation}
Here $z$ is a radial direction, and $\ell _\ads$ is an AdS radius.

\medskip

According to the Ryu-Takayanagi formula,  the holographic entanglement entropy of a half plane
 is given by 
\begin{equation}
S_A=\frac{V_{d-2}}{4G_{\rm N}}\int_{\varepsilon}^\infty \! \dd z \, \left( \frac{\ell _\ads}{z}\right)^{d-1}
=\frac{V_{d-2}\ell _\ads^{d-1}}{4G_{\rm N}(d-2)}\cdot \frac{1}{\varepsilon ^{d-2}}
\label{eeads}
\end{equation}
where $G_{\rm N}$ is Newton's constant, 
and $\varepsilon$ is a UV cutoff. This is the result in  the AdS/CFT correspondence.

\medskip

Performing the double Wick rotation \eqref{wick} while Newton's constant $G_{\rm N}$ and 
the UV cutoff $\varepsilon$ are held fixed, 
the above entanglement entropy \eqref{eeads} becomes
\begin{equation}
S_A=(-i)^{d-1}\frac{V_{d-2}\ell _\ds^{d-1}}{4G_{\rm N}(d-2)}\cdot \frac{1}{\varepsilon ^{d-2}}\,.
\label{eeds}
\end{equation}
In the AdS case, the  extremal  surface is $0\leq z< \infty$ at $x_1=0$.
After the double Wick rotation, the extremal surface is given by 
\begin{equation}
0 \leq \eta <i \infty \,.
\end{equation}
The extremal surface in dS is not real-valued but {\it complex}-valued.
The idea of complex surfaces has also appeared in the AdS case \cite{FM}.

\subsection{Extremal surfaces in asymptotically dS}
In the previous subsection, we found the extremal surface in Poincar\'{e} dS using the double Wick rotation.
We comment on  extremal  surfaces in a more general set of asymptotically dS.

\medskip

To define extremal surfaces in asymptotically dS, we need to find a double Wick rotation between 
the asymptotically dS and the corresponding asymptotically EAdS.
One Wick rotation is 
\begin{equation}
\ell _{\ds} \to i \ell _{\ads}
\end{equation}
to make the cosmological constant positive.
A second analytic continuation is concerned with a time coordinate in dS.

\medskip

Our proposal is that the holographic entanglement entropy in the dS/CFT correspondence is defined as 
\begin{equation}
S_A:=\frac{\text{Area}_\ds}{4G_{\rm N}}\,.
\label{asym-ds}
\end{equation}
Here Area${}_\ds$ is the area of the ``extremal surfaces'' in asymptotically dS and is defined as follows.
First of all, we find extremal surfaces in asymptotically EAdS.
Next, performing the  double Wick rotation of the extremal surfaces in asymptotically EAdS,
we define ``extremal surfaces'' Area${}_\ds$ in asymptotically dS.
As in the previous subsection, the extremal surfaces in dS are {\it complex}-valued 
in general although the extremal surfaces in AdS are real-valued.
The holographic entanglement entropy \eqref{asym-ds} is uniquely defined 
by using the extremal surfaces in asymptotically EAdS.

\section{Comparison with A Toy CFT Model}

Since the conformal field theory (CFT) dual to Einstein gravity on dS is not  known yet, 
we analyze the free $Sp(N)$ model as a toy model.
Since the $Sp(N)$ model is the holographic dual of Vasiliev's higher-spin gauge theory on dS, 
we can quantitatively compare such results only with Vasiliev's higher-spin gauge theory, not with Einstein gravity.
However, it is natural to expect that their basic qualitative behaviours do not change between these two theories. 

\medskip

The free $Sp(N)$ model on a Euclidean space with the metric $g_{ij}$ is defined  by the action 
\begin{equation}
I=\int \! \dd^d x \,  \sqrt{g }\, \Omega _{ab} g^{ij}\partial_i \chi ^a  \partial_j \chi ^b \,, 
\qquad \Omega _{ab}=\left(
    \begin{array}{cc}
      0 & 1_{N/2\times N/2}  \\
      -1_{N/2\times N/2} & 0 
    \end{array}
  \right)\,,
\end{equation}
where $\chi ^a$ $(a=1,\cdots ,N)$ are anticommuting scalars, and $N$ is an even integer \cite{spcft}.
By introducing 
\begin{equation}
\eta^a= \chi ^a+i\chi ^{a+\frac{N}{2}} \,, \quad 
\bar{\eta}^a=-i \chi ^a-\chi ^{a+\frac{N}{2}}  \qquad \left( a=1,\cdots , \frac{N}{2} \right)\,,
\end{equation}
the action is rewritten as  
\begin{equation}
I=\int \! \dd^d x \, \sqrt{g} \, g^{ij} \partial_i \bar{\eta} \, \partial_j \eta \,.
\end{equation}

\medskip

Let us calculate the
``entanglement entropy'' in the free $Sp(N)$ model on $\mathbb{R}^d$ $(g_{ij}=\delta_{ij})$.
We divide the $x_1$ slice of  $\mathbb{R}^d$ into two regions $A$ and $B$, and define the 
entanglement entropy $S_A$ as,
\begin{equation}
S_A:=-{\rm Tr}_A \, \rho_A \log \rho_A 
\end{equation}
Here the  reduced density matrix $\rho _A$ is defined as $\rho _A={\rm Tr}_B\, \rho$ 
by using the total density matrix $\rho$.
For simplicity, we take $x_2\geq 0$ as the subsystem $A$.
By using the replica trick, the entanglement entropy can be expressed as
\begin{equation}
S_A=-\lim _{n\to 1}\frac{\partial }{\partial (1/n)}\left( \log Z_{\mathbb{R}^2/\mathbb{Z}_n \times \mathbb{R}^{d-2}}
-\frac{1}{n}\log Z_{\mathbb{R}^d}\right) \,,
\end{equation}
where $Z_{\mathbb{R}^2/\mathbb{Z}_n \times \mathbb{R}^{d-2}}$ and $Z_{\mathbb{R}^d}$ are
partition functions on $\mathbb{R}^2/\mathbb{Z}_n \times \mathbb{R}^{d-2}$ and $\mathbb{R}^d$, respectively. 
The logarithm of the partition function is evaluated as 
\begin{equation}
\log Z_{\mathbb{R}^d}=\log \int \! \pd \bar{\eta}\,  \pd \eta \,  {\rm e}^{-I}=
NV_d \log \int \! \frac{\dd^d k}{(2\pi)^d} \log k^2\,,
\end{equation}
where $V_d$ is the volume of $\mathbb{R}^d$.
Note that this result is  minus that of standard field theories.
It comes from the statics of the fields.
Since $\log Z_{\mathbb{R}^2/\mathbb{Z}_n \times \mathbb{R}^{d-2}}$ is also 
minus that of the standard field theories, the entanglement entropy is given by  
\begin{equation}
S_A=-\frac{NV_{d-2}}{6(d-2)(4\pi)^{\frac{d-2}{2}}} \cdot \frac{1}{\varepsilon ^{d-2}} \,,
\label{ee}
\end{equation}
where $V_{d-2}$ is a $(d-2)$-dimensional infinite volume, and $\varepsilon$ is a UV cutoff.
The entanglement entropy \eqref{ee} is minus that of  standard field theories.
We can also obtain similar results for arbitrary subsystems on $\mathbb{R}^d$ or other curved spaces.

\medskip

The result \eqref{ee} holds for any dimensions.
However note that it is known that the duality between Vasiliev's higher-spin gauge theory 
and the $Sp(N)$ model holds only when $d=3$.

\section{Conclusion and discussion}

In this paper,  we have discussed the entanglement entropy in the dS/CFT correspondence.
We have proposed the holographic entanglement formula for Einstein gravity on dS and 
asymptotically dS via the double Wick rotation.
In our proposal we found extremal surfaces which extend in complex-valued coordinate spacetime.
The holographic entanglement entropy behaves as $S_A\propto (-i)^{d-1}$ in dS${}_{d+1}$ \eqref{eeds}.

\medskip

To check that our proposal works, we have calculated the entanglement entropy in the 
free $Sp(N)$ model and compared it with our proposal. 
We have found that the entanglement entropy is given by minus that of  standard field theories \eqref{ee}.
This result may suggest that the dS$_{d+1}$/CFT$_d$ correspondence makes senses 
only when $d \in 4\mathbb{Z}-1$.
Note that the most interesting case, the dS${}_4$/CFT${}_3$ correspondence, is included, while the most simple 
case, the dS${}_3$/CFT${}_2$ correspondence, is excluded.
This is consistent with results in subsection 5.2 in \cite{nonG}.

\medskip

Our proposal has been checked only in the simple case, where the subsystem is the half plane.
We need to check our proposal in more nontrivial setups with circular or 
some other shaped entanglement surfaces for example.
However, it is expected that our proposal holds in any entanglement surfaces because 
the factor $(-i)^{d-1}$ appears in Einstein gravity via the double Wick rotation \eqref{eeds}, and 
 the minus sign appears in the $Sp(N)$ model \eqref{ee}.
It is interesting to apply our proposal to Schwarzschild dS.
Naively, it is expected that the holographic entanglement entropy is a sum of 
that of  pure dS  and the Schwarzschild BH entropy.

\medskip

Finally, we comment on the negativity of the entanglement entropy \eqref{ee} in the free $Sp(N)$ model.
In standard field theories, entanglement entropies are positive definite.
In contrast, our result \eqref{ee} is negative definite.
The negativity  comes from the fact  that the scalars of the $Sp(N)$ model are anticommuting, 
and implies that the inner products of the Hilbert space are not positive definite.
This negativity might be a key ingredient of the dS/CFT correspondence.

\section*{Acknowledgments}

We would like to thank Tadashi Takayanagi for useful comments and discussions. 
We also thank Tomoki Nosaka for useful comments on the draft.
The work  is supported by a Grant-in-Aid for Japan Society for
the Promotion of Science (JSPS) Fellows No.26$\cdot$1300. 

\subsubsection*{Note added:}

While this work was in progress, the article \cite{Narayan} appeared in the arXiv.
Our results in section 2 are overlapped with \cite{Narayan}.

\end{document}